\documentclass[prd,showpacs,onecolumn]{revtex4}
\usepackage{psfrag,graphics,dcolumn,bm}

\begin{document}

\newcommand{\beq}{\begin{equation}}
\newcommand{\eeq}{\end{equation}}
\newcommand{\beqa}{\begin{eqnarray}}
\newcommand{\eeqa}{\end{eqnarray}}

\newcommand{\lexp}{\mathop{\langle}}
\newcommand{\rexp}{\mathop{\rangle}}
\newcommand{\rexpc}{\mathop{\rangle_c}}
\newcommand{\mnras}{MNRAS}
\newcommand{\apjl}{ApJ}
\newcommand{\apss}{Ap\&SS}
\newcommand{\apjs}{ApJS}
\newcommand{\aap}{A\&A}

\def\bi#1{\hbox{\boldmath{$#1$}}}

\title{A comparison of cosmological Boltzmann codes:
are we ready for high precision cosmology?} 

\author{Uro\v s Seljak}
\affiliation{Department of Physics,
Jadwin Hall, Princeton University, Princeton, NJ 08544 }

\author{Naoshi Sugiyama}
\affiliation{Division of Theoretical Astrophysics, National Astronomical
Observatory Japan, Tokyo, 181-8588, Japan}
\author{Martin White}
\affiliation{Departments of Physics and Astronomy, University of California,
Berkeley, CA 94720}
\author{Matias Zaldarriaga}
\affiliation{Department of Astronomy and Department of Physics,
Harvard University, Cambridge, MA 02138}
\date{\today}

\begin{abstract}
We compare three independent, cosmological linear perturbation theory codes
to asses the level of agreement between them and to improve upon it by
investigating the sources of discrepancy.  By eliminating the major sources
of numerical instability the final level of agreement between the codes was
improved by an order of magnitude.
The relative error is now below $10^{-3}$ for the dark matter power spectrum.
For the cosmic microwave background anisotropies the agreement is below the
sampling variance up to $l=3000$, with close to $10^{-3}$ accuracy reached
over most of this range of scales.
The same level of agreement is also achieved for the polarization spectrum
and the temperature-polarization cross-spectrum. 
Linear perturbation theory codes are thus well prepared for the present and
upcoming high precision cosmological observations.
\end{abstract}



\maketitle
\section{Introduction}

Since the first detection of anisotropies in the cosmic microwave background
(CMB) over a decade ago \cite{1992ApJ...396L...1S}
progress has been steady and rapid.
These measurements are in astonishing agreement with theoretical predictions
of adiabatic cold dark matter models which have been refined over the last
few years.
The standard model emerging from these measurements suggests that we live
in a spatially flat universe dominated by dark energy and dark matter,
with a small amount of baryons and a spectrum of primordial fluctuations
that is close to scale invariant.  
This picture has been given its most dramatic confirmation by the 
recent WMAP results \cite{2003astro.ph..2207B,2003astro.ph..2209S},
which have reached percent level accuracy on degree scales, in
combination with small-scale anisotropy measurements
\cite{2002astro.ph..5384M,2002astro.ph.12289K}.

While the current observational situation is already impressive, future
observations are even more promising.
The {\sl Planck\/} satellite and several ground based small scale CMB
experiments
(APEX, 
SPT, 
ACT) 
will reach sub-percent accuracy on scales above $10'$.
A possible next generation CMB satellite dedicated to polarization could
measure polarization to a comparable accuracy, as well as measure the
projected dark matter potential using the lensing effect on CMB.
High precision cosmology will also be achieved with other data sets, most
notably large scale structure (LSS) and supernovae, both of which can
supplement the information from the CMB to break the degeneracies 
\cite{1998ApJ...506..495W,1999ApJ...518....2E,1999ApJ...514L..65H,2003astro.ph..2112M,2002astro.ph.10217L}. 
Current constraints from galaxy clustering, weak lensing and the Ly-$\alpha$ 
forest are limited by either statistics or systematic effects in the analysis
and have not yet reached percent level precision.
However, with better data and more work on systematics both of these aspects
should improve dramatically.
Comparison between the different probes will also provide additional
cross-checks on the systematics.

High precision cosmological observations are of course pointless if they
are not matched by theoretical predictions.
The CMB and, to a lesser extent, LSS are unique in that they are sensitive
to perturbations in linear regime.
In this regime the evolution equations can in principle be solved to arbitrary
precision and are thus limited only by the accuracy of the linear
approximation itself.
In practice the computational task is not quite so simple for various reasons:
the evolution equations are complicated, the solutions are highly oscillatory
and thus susceptible to numerical errors, the equations can be stiff and
require different treatments in different regimes etc. 

In this era of high precision cosmology it is worth revisiting the status of
the theoretical calculations as well.
The last of such comparison was performed almost a decade ago
\cite{1995PhRvD..52.5498H}.
Informal comparisons between the different groups at the time led to a
nominally stated accuracy of 1\% for codes such as
CMBFAST \cite{1996ApJ...469..437S}. 
At the time both the CMB and LSS measurements were much less precise and
in the case of CMB limited to large scales, where sampling variance limits
the required accuracy. 
Thus 1\% precision was more than sufficient for measurements then. 
Today, systematic errors at 1\% level are already comparable to the statistical
precision of the current observations such as WMAP and will certainly not
suffice for the next generation of CMB experiments.
Moreover, a decade ago the standard cosmological model was a flat cold dark
matter (CDM) model with $\Omega_m=1$ and there were only limited comparisons
performed for the currently favored model with significant dark
energy/cosmological constant or reionization optical depth.
The goal of this paper is to revisit the accuracy of the current linear
perturbation theory codes.  

It is useful to provide here some history on the development of relativistic 
perturbation theory and Boltzmann codes.
Initial work on perturbation theory, including the classification of
perturbations into scalar, vector and tensor, was done by
Lifshitz \cite{1946jphyussr10.116.L}.
Later papers clarified the gauge issues for scalar modes
\cite{1980PhRvD..22.1882B,1984ptp...78..1K}.
The main ingredients for computing the CMB spectrum were put in place already
by early seventies \cite{1967ApJ...147...73S,RS68,1968ApJ...151..459S,1970ApJ...162..815P,1970Ap&SS...7....3S, 1978SvA....22..523D},
in those days still without cold dark matter (CDM). 
Work in the eighties introduced CDM and several computational advancements,
such as the use of the multipole moment hierarchy to solve the equations
for photon distribution function and the introduction of polarization
\cite{1981ApJ...243...14W,1984ApJ...285L..39V,1984ApJ...285L..45B,1987MNRAS.226..655B}. 
Code development became an active area of research in the early nineties and
there were several codes in addition to the ones mentioned above developed
around that time
\cite{1989ApJS...71....1H,1993PhRvL..70.2224D,1994A&A...287..693S}. 
As new cosmological models or parameters were introduced the corresponding
CMB spectrum was calculated, such was the case for open models 
\cite{1983ApJ...273....2W,1991PThPh..85.1023G},
tensors modes \cite{1993PhRvL..71..324C}
and massive neutrinos \cite{1994santanderS,1996ApJ...467...10D}. 
A new method to compute the anisotropies based on line of sight integration
was introduced in 1996 \cite{1996ApJ...469..437S}.
The resulting public domain code named CMBFAST was roughly two orders of
magnitude faster than the traditional Boltzmann codes.
The main subsequent developments were the improved treatment of polarization
including $E$ and $B$ modes
\cite{1997PhRvD..55.1830Z,1997PhRvD..55.7368K,1997PhRvD..56..596H}, 
inclusion of lensing effect on the CMB
\cite{1989MNRAS.239..195C,1996ApJ...463....1S,1998PhRvD..58b3003Z,2000PhRvD..62d3007H},
spatially closed models
\cite{1996ApJ...459..415W,2000ApJ...538..473L,2000ApJS..129..431Z},
improvements in the recombination calculation
\cite{1999ApJ...523L...1S,2000ApJS..128..407S}
and introduction of additional cosmological parameters, such as dark
energy/quintessence \cite{1997PhRvD..56.4439T,1998PhRvL..80.1582C}.

The principal guidelines in deciding which codes to include in the current
comparison were independence and accuracy.
While there was a lot of code development activity after COBE, most of the
codes were not being updated after CMBFAST was made public. 
Two exceptions to this are the codes developed by
N.~Sugiyama \cite{1992PThPh..88..803S,1995ApJS..100..281S}, hereafter NS, and
M.~White \cite{1996ApJ...459..415W,1997PhRvD..56..596H,1998PhRvD..57.3290H},
hereafter MW, both of which are included in the current comparison.
These two codes are completely independent of CMBFAST and are traditional
Boltzmann codes without the line of sight integration.
NS code is based on gauge invariant formalism, while MW code and CMBFAST 
use synchronous gauge formalism.
Other, more recent codes, such as CAMB \cite{2000ApJ...538..473L} and
CMBEASY \cite{2003astro.ph..2138D}, originally started as translations
of CMBFAST into f90 and C++, respectively, and are thus not independent.
There was a lot of subsequent work put into these codes later, so the
extent to which the possible numerical errors in CMBFAST are also present in
these codes is unclear and we do not explore it in this paper.
We also do not use the COSMICS package \cite{1995astro.ph..6070B} in this
comparison. 
CMBFAST Boltzmann evolution equations originate from COSMICS and are
thus not independent.
At the time of CMBFAST development the two codes were extensively compared,
but many of subsequent developments (polarization, non-flat geometries,
lensing, dark energy) were not included in the COSMICS package.

The goal of this paper is to test the numerical accuracy of the linear
perturbation theory codes, which are used extensively in the parameter 
determinations.
When we started the project the initial agreement was no worse than 1-2\%,
consistent with the stated accuracy.
We will show below that the final agreement is much more impressive than
that and is essentially sampling variance limited up to the highest
multipole moment we used in comparison ($l=3000$). 
This is not to say that the theoretical predictions are this accurate,
since the physics used in the codes is the same and there could be
additional effects not included in any of the current versions.
However, the numerical approximations, which are present in all of the
codes, appear to be under control and do not lead to systematic errors
of significance for the current and next generation of experiments.

\section{Comparison of results}

In the current code comparison we limit ourselves to the simplest model
with a cosmological constant.
Even though the model was chosen prior to recent WMAP results it is in
fact very close to their best fit model.
Our standard model has $\Omega_{CDM}=0.3$, $\Omega_b=0.04$,
$\Omega_{\Lambda}=0.66$ and $H_0=70$km/s/Mpc.
We assume a scale invariant $n=1$ primordial power spectrum with and without
reionization (since the results are for the most part 
unchanged in the two cases we will
only show those without reionization in the following).
We do not include gravitational lensing effect in the current comparison,
since it is not implemented in all of the codes (efforts to verify the
lensing code accuracy in CMBFAST are currently underway). 
For the same reason we also do not include the tensors in our comparison.
Accuracy of the tensor calculation is unlikely to be critical for the
current or future generation of experiments, since tensors are already
known to be subdominant and only contribute on large scales, where
the sampling variance errors are large.
Similar arguments also apply to the massive neutrinos or more general forms
of dark energy, which are thus not explored in more detail here.
In all of the comparisons we used the same recombination outputs.
We have found some small differences between the different implementations
of RECFAST \cite{1999ApJ...523L...1S}, but these appear not to be important
at the current level of precision. 

The required accuracy depends on the scale one is probing and the information
one is extracting from the power spectra. 
For a given $C_l$ there are $2l+1$ independent multipole moments in the sky
and the relative error on it will be roughly given by $1/\sqrt{l}$.
However, the ultimate goal is not the spectrum $C_l$, but a small set of
cosmological parameters, so one must combine $C_l$'s together.
If the errors in the calculation of $C_l$ are correlated then one needs a
more stringent accuracy criterion.
In the limit of only one parameter being determined from the data
(for example, the overall amplitude of the spectrum assuming its shape
is known) the number of modes up to $l$ is $l^2$ and the required theoretical
precision is $\sqrt{2}/l$.
This theoretical limit is not reached in practice, since there is always
more than one parameter determined from the CMB data and since the sky
coverage is always less than unity
(due to the finite sky coverage or galactic contamination).
To account for this we will roughly double this limit, so that we assume
the required precision at a given $l$ is determined by 
\begin{equation}
   {\delta C_l \over C_l}={3 \over l}.
\label{cv}
\end{equation}
This corresponds to 0.1\% accuracy at $l=3000$, the maximum $l$ used in 
comparison here. Note that for $l<30$ the required accuracy is only 10\%
and there is thus little point in attempting to achieve very high 
accuracy on large scales.

\begin{figure}
\begin{center}
\resizebox{5in}{!}{\includegraphics{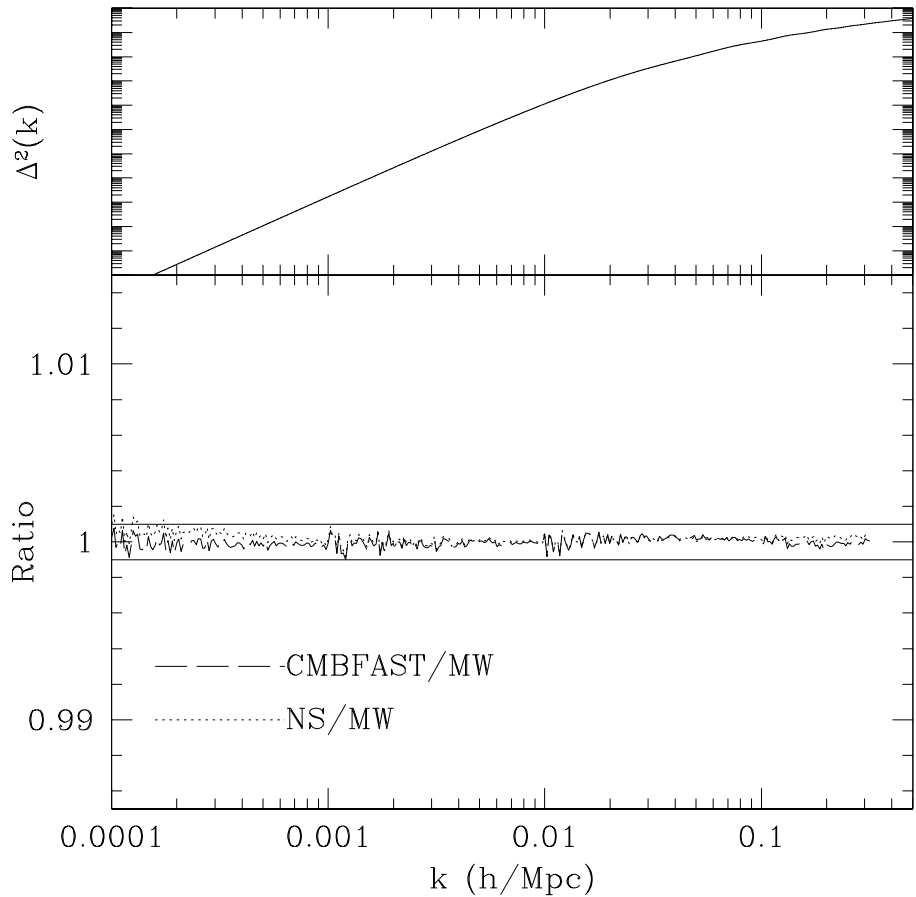}}
\end{center}
\caption{Dark matter power spectrum for the 3 codes (top) and 
ratios between them (bottom). Also shown are $1\pm 0.1$\% horizontal lines.  
The relative errors are below 0.1\%.}
\label{fig1}
\end{figure}

While for the CMB the sampling variance always limits the required theoretical 
precision, this is less of an issue for the 3-d matter power spectrum. 
Fortunately, the matter power spectrum is also much easier to compute with 
high accuracy.
Figure \ref{fig1} shows the comparison between the matter power spectra among
the three codes.
We have assumed the same initial conditions in all the codes, so the
comparison of the transfer functions at the end tests the accuracy of
relating the primordial spectrum of fluctuations to the final matter power
spectrum, both in normalization and shape.
We see that the agreement is remarkable, at least at the level of $10^{-3}$.
This is comparable or better than the accuracy in the CMB, so any matter
power spectrum normalization from the CMB (such as $\sigma_8$) is limited
by the accuracy in the CMB spectra.
Computing the dark matter (as well as baryon or massive neutrino) power
spectra is thus essentially exact for the current purposes. 
It is easy to understand why the dark matter transfer function can be
computed so accurately.
The evolution equation for the dark matter is a simple second order
differential equation,
its solutions are smooth and have a simple power law (or logarithmic) growth
both in radiation and matter domination epochs.
Dark matter exhibits no oscillatory behaviour and only couples to gravity.
As a result its evolution can be computed numerically to an exquisite
precision.

We turn next to the CMB comparison.
In general, high accuracy of CMB anisotropies is much more difficult to
achieve than that of the dark matter power spectrum.
There are several reasons for this: 

1) Before recombination, the evolution of baryons and photons is tightly
coupled due to the high probability of a photon scattering off an electron.
This leads to a stiff system of differential equations and a special treatment
must be used before recombination, switching to the regular one at later
times when the mean free path to Thomson scattering increases. 

2) As is well known, the CMB spectra have a lot of structure with prominent
acoustic peaks, unlike the dark matter where the spectrum is a slowly varying
function of scale. The structure is even sharper for polarization spectrum
and polarization-temperature cross-correlation (where the spectrum can be
positive of negative). 
The phases of the acoustic peaks depend sensitively on the numerical accuracy.
They also depend sensitively on the recombination history, which must be
computed very accurately. 

3) The time dependence of the multipole moments is highly oscillatory and
requires fine time sampling. To obtain a $C_l$ at a given $l$ one must
integrate over all the Fourier modes $k$.  This $k$-mode dependence is also
highly oscillatory and again requires fine sampling to achieve a sufficient
accuracy.  For traditional Boltzmann codes this can be computationally
expensive, so approximations have been developed to reduce the number of
evaluations.  This is in principle avoided in the line-of-sight integration
approach used in CMBFAST, which however introduces its own approximations.
Among these are the time sampling of the sources, treatment of reionization,
$l$ sampling, cutoff in the photon and neutrino hierarchy etc.
At the time of CMBFAST first release the main goal was to reduce the
computational time while still maintaining 1\% accuracy.  The approximation
criteria were often chosen aggresively to reduce the run time. 
We have found that many of these approximations can be significantly improved
in accuracy if original criteria are made slightly more conservative, without
a significant increase in the run time. 

\begin{figure}
\begin{center}
\resizebox{5in}{!}{\includegraphics{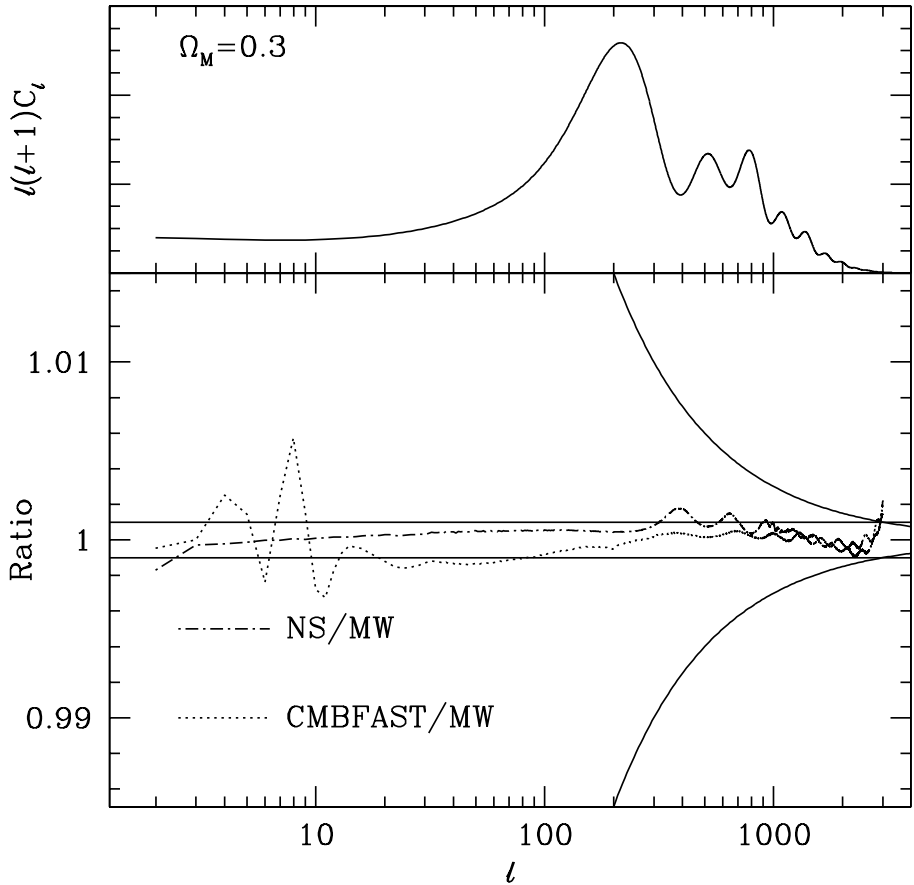}}
\end{center}
\caption{$C_l^{TT}$ for the 3 codes (top) and 
ratios between them (bottom). Also 
shown is the sampling variance limit $1\pm 3/l$ and 
$1\pm 0.1$\% horizontal lines.}
\label{fig2}
\end{figure}

Figure \ref{fig2} shows the ratios between the codes for the
temperature spectrum $C_l^{TT}$.
We also show the cosmic variance error (equation \ref{cv}) and $\pm$0.1\% error
lines.
We see that the agreement is well within the cosmic variance limits and
close to 0.1\% for almost all $l$.
The only exception is around $l\sim 10$, where there is a somewhat larger
error of up to 0.5\%.
This is caused by the line of sight integration method as implemented in
CMBFAST, where one uses integration by parts to rewrite the sources into a
single term that multiplies the spherical Bessel functions.
This form requires very precise cancellations in the integrals over the
visibility function on large scales.
The error is however harmless, since it is two orders of magnitude smaller
than the sampling variance.
The agreement between NS and MW is equally remarkable and even better than 
MW/CMBFAST on large scales.

\begin{figure}
\begin{center}
\resizebox{5in}{!}{\includegraphics{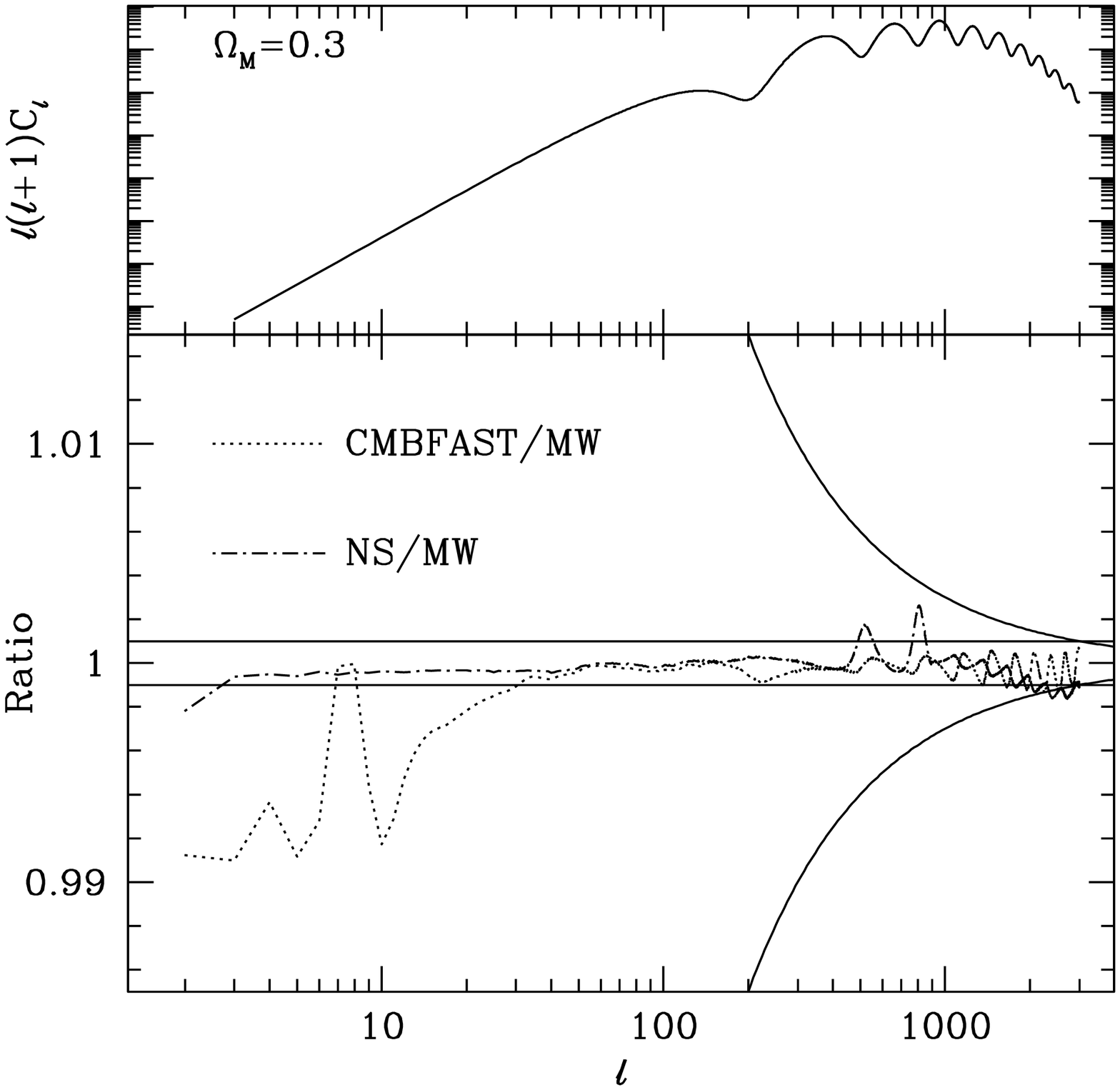}}
\end{center}
\caption{Same as figure \ref{fig2} for $C_l^{EE}$.
}
\label{fig3}
\end{figure}

Figure \ref{fig3} shows the same comparison for $C_l^{EE}$, the E-type
polarization power spectrum.
The agreement is roughly at the same level as for $C_l^{TT}$, close to
0.1\% across all the scales. The exception again is CMBFAST at $l<20$, 
where the error can be up to 1\%, caused by imperfect cancellations
in the line of sight integration
over the recombination epoch at
$z\sim 1100$. This discrepancy is not a real 
problem, since the comparisons here are for no reionization model and 
even a small amount of reionization increases the polarization 
power at low $l$ with a contribution from $z<z_{\rm reion}\sim 10-20$. 
Our reionization model comparisons show a better agreement. 
In any case, in this regime even a 1\% error 
is a factor of at least 10 lower than the sampling variance and  
thus irrelevant. 

\begin{figure}
\begin{center}
\resizebox{5in}{!}{\includegraphics{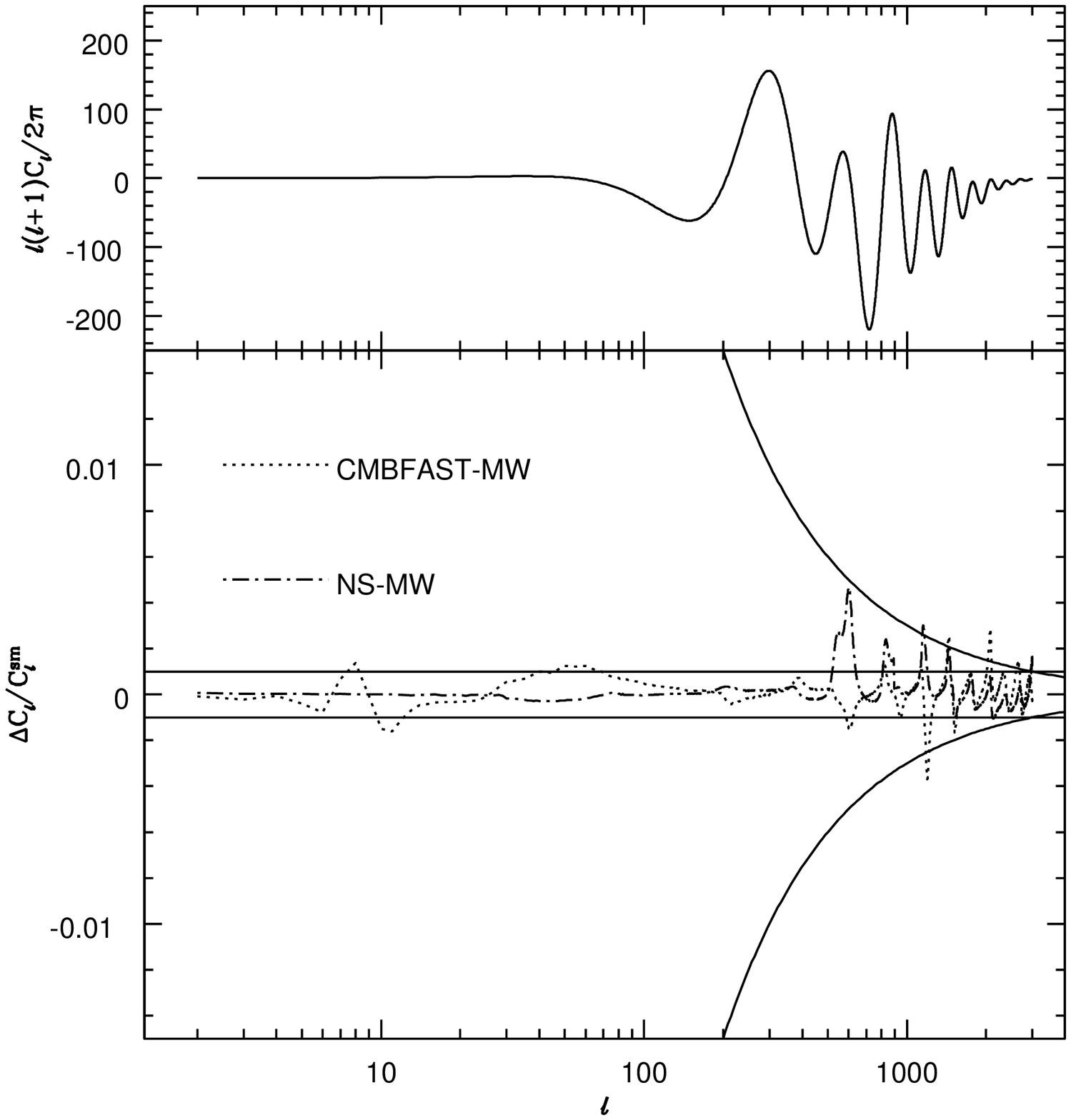}}
\end{center}
\caption{Same as figure \ref{fig2} for $C_l^{TE}$.
At zero crossings of $C_l^{TE}$ the relative error 
is ill-defined, so we compare to 
a smoothed version, where the smoothing is $\Delta l =50$.  
The plotted sampling variance limit $1\pm 3/l$ is a lower 
limit to the actual sampling variance, as discussed in the text.
}
\label{fig4}
\end{figure}

Finally, figure \ref{fig4} shows the temperature-polarization
cross-correlation $C_l^{TE}$.
The relative error is ill-defined at zero crossings of $C_l^{TE}$.
For this reason we compare to a smoothed version of $C_l^{TE}$, 
smoothing over $\Delta l =50$. The agreement is again very 
good, no worse than for $C_l^{TT}$ or $C_l^{EE}$.
Note that the sampling variance for $C_l^{TE}$ at a fixed $l$ is given by 
$\delta C_l^{TE}/C_l^{TE}\sim \sqrt{2/[l(1+C_l^{TT}C_l^{EE}/(C_l^{TE})^2)]}$
\cite{1997ApJ...482....6S}, which is always larger than the corresponding 
limits for $C_l^{TT}$ and $C_l^{EE}$,   
so our plotted sampling variance limit is a conservative lower limit. 
We find a similar level of agreement when comparing 
the absolute errors without smoothing, which are also at the level of  0.1\%.

\section{Conclusions}

We have performed a comparison of 3 current high accuracy linear perturbation
theory codes.
The initial agreement was at 1\% level, while the final one was at 0.1\% level,
an order of magnitude improvement. 
The same 0.1\% accuracy is also found for polarization and its
cross-correlation with temperature.
For the dark matter power spectrum the agreement is also at 0.1\% level or
even better.
It seems unlikely that we will ever need better accuracy than this both for
the CMB and for the matter power spectrum.
The theoretical predictions of the CMB and matter power spectra are thus well
under control, at least for the codes and models 
used in the current comparison.
We note that the modifications needed to upgrade the CMBFAST code to
0.1\% accuracy have been implemented in version 4.2, which is available from
www.cmbfast.org. As a caveat we note that the open/closed model implemenation 
remains at 1\% level and that the accuracies of lensing, massive neutrinos 
and dark energy remain to be explicitly verified. Some of these 
comparisons are currently in progress.

The main remaining concern are the physical assumptions that enter into the
calculations.
These have been scrutinized by many workers over the past decade
(see e.g. \cite{1995PhRvD..52.5498H}),
which gives us some confidence that there cannot be too many physical
processes that have been overlooked by now.
The principal concern at the moment is the accuracy of the recombination
calculation.
The original treatment \cite{1968ApJ...153....1P,1968ZhETF..55..278Z}
has been revisited in \cite{1999ApJ...523L...1S}.
It was found that the original work by Peebles was remarkably accurate,
but there were some improvements at the level of a few percent.
For example, it was shown that HeI recombination cannot be well described
by the Saha equation approximation and that the Boltzmann equilibrium
assumption for the higher levels of hydrogen was not sufficiently accurate.
The latter can be approximated by a fudge factor added to the previous
treatment. 
These changes lead to a few percent differences in the CMB spectrum and
were implemented into the RECFAST routine \cite{1999ApJ...523L...1S}.
While we have no reason to suspect the accuracy of these calculations
it is also not obvious that it is at the 0.1\% which will be needed for
the next generation of CMB experiments.
Thus the accuracy of the CMB spectrum calculations may well be limited by the
treatment of the recombination and it would thus be useful to revisit this
issue to asses the level of remaining uncertainty.
If this proves to be larger than the upcoming experimental sensitivity then
a possible approach would be to parametrize the uncertainty in the 
recombination physics and to reduce the uncertainty directly from the
observations.
As a simple example, if hydrogen recombination rate is uncertain then one
could treat the fudge factor mentioned above as a free parameter that one
could determine directly from CMB observations. 
It is an open question at present how uncertainty in the physics governing
recombination feeds into measurements or reconstructions which rely on the
CMB damping tail.

The new generation of the CMB experiments under construction or planning
will achieve a sub-percent accuracy on several cosmological parameters.
Of special importance are the parameters related to the shape of the
primordial power spectrum, which should be determined with exquisite
precision.
Such measurements will allow high precision tests of early universe models
such as various models of inflation.
This can however only be achieved if theoretical predictions match the
observations in accuracy.
It is comforting to know that the numerical precision of linear calculations
is not among the worries for the future of high precision cosmology.

Acknowledgements: we acknowledge the hospitality of the
Institute for Theoretical Physics at Santa Barbara. US 
is supported by 
Packard Foundation, Sloan Foundation 
NASA NAG5-1993 and NSF CAREER-0132953.
NS is supported by the Alexander von Humboldt foundation and 
Japanese Grant-in-Aid for Science Research Fund of the Ministry of
Education, No.14340290. MW is supported by NSF and NASA. MZ is 
supported by Packard Foundation and NSF.

\bibliography{apjmnemonic,cosmo,cosmo_preprints}   

\end{document}